\documentclass[12pt]{article}

\usepackage{endnotes}

\usepackage{graphicx}  
\usepackage{amsmath}   
\usepackage{amssymb}   

\usepackage[active]{srcltx}

\def\eq#1{{Eq.~(\ref{#1})}}

  \title{One hundred years of General Relativity: Summary, Status and Prospects}
  \author{T. Padmanabhan\\
  IUCAA, Pune University Campus,\\
  Ganeshkhind, Pune- 411 007.\\
  {\small {email: paddy@iucaa.ernet.in}}
  }

  \date{}  

\begin{document}
  
  \maketitle
  
 \begin{abstract}
General Relativity (GR) revolutionized  the way we we thought about gravity.  After briefly describing  the key
successes of GR and  its impact,  I will discuss the major conceptual challenges it faces today.
I conclude by  outlining the prospective future directions of development,  which hold the promise
of deepening our understanding of the nature of gravity.
\end{abstract}

 

The general theory of relativity, produced by Einstein in 1915, has received innumerable accolades, of which my personal favourite is the sentence which occurs in Volume II of the Course of Theoretical Physics by Landau and Liftshitz: ``\textit{It ... represents probably the most beautiful of all existing physical theories}.'' What is remarkable is not the quote but the fact that the authors --- well known for the terse style in which not a single word is wasted in empty ceremonies  --- were forced to pause and make this comment!  

General relativity interprets the gravitational effects  as arising due to the curvature of spacetime. 
Like all profound ideas, this one  --- viz., gravity \textit{is} the curvature of the spacetime --- is  obvious with hindsight! Start from the elementary fact that 
 two bodies of different masses, dropped near the earth's surface, will  fall downwards  with the same acceleration  $g$ and  hit the ground simultaneously. Consider now the same bodies, released gently inside an elevator which is moving upwards with the acceleration $g$ in interstellar space. The bodies will stay at the locations where they were released, because there is no force acting on them. But the floor of the elevator will come up and hit them \textit{simultaneously}, making an observer in the elevator believe that the two masses were moving downwards with the same acceleration $g$. So the observer cannot distinguish (locally) between a gravitational field and an accelerated elevator  by this experiment. Einstein generalized this idea as a \textit{Principle of Equivalence} and postulated that \textit{no physical phenomenon} can distinguish between the two situations. 

Next, recall a result from special relativity, viz. that when a clock moves a distance $d\textbf{x}=\textbf{v} dt$ in a time interval $dt$ (measured by \textit{your} clock), it will register a lapse of time $d\tau$ where $c^2d\tau^2=c^2 dt^2-d\textbf{x}^2$.  So if we attach a clock to the floor of an accelerated elevator, its resulting motion will slow it down. Principle of Equivalence demands the clock should also slow down in an equivalent gravitational field, in order to prevent us from distinguishing an accelerated elevator from a gravitational field using clocks. So the rate of flow of time   in a gravitational field must be affected by the gravitational potential. 

One can show that this requires  modifying the Pythagorean addition rule $c^2d\tau^2 = c^2dt^2 - d\textbf{x}^2$ to a more complicated quadratic expression, which reduces to 
\begin{equation}
c^2d\tau^2= [1+(2\phi/c^2)]c^2dt^2 - d\textbf{x}^2
\end{equation} 
in a weak gravitational potential $\phi$.  Such an addition of intervals is mathematically equivalent to assuming that the spacetime is curved. Voila! The presence of a gravitational field is equivalent to the curvature of spacetime. It is a long way from here to obtain the full theory of GR, but the rest, as they say in chess, is essentially a matter of precise technique. The genius part is over.\endnote{Einstein struggled to find the correct field equations because he was trying to keep the source as the stress tensor. An alternative way of generalizing the Newtonian law $\nabla^2\phi\propto\rho$ is as follows (see e.g., p. 259 of \cite{key7}). One first notes that: (i) The energy density $\rho=T_{ab}u^au^b$ is foliation/observer dependent where $u^i$ is the four velocity of an observer. (ii) Since $g_{ab}$ plays the role of $\phi/c^2$, the left hand side $\nabla^2\phi$ should come from the curvature tensor and should depend on the observer (since the right hand side does). (iii) It is perfectly acceptable for the left hand side \textit{not} to have second time derivatives of the metric in the rest frame of the observer. 

To obtain something analogous to \textit{spatial} second derivatives, one begins by projecting the indices of $R_{abcd}$ to the space orthogonal to $u^i$, using the projection tensor $P^i_j=\delta^i_j-u^iu_j$, obtaining the tensor
$\mathcal{R}_{ijkl}\equiv P^a_iP^b_jP^c_kP^d_l R_{abcd}$. The only scalar we can get from $\mathcal{R}_{ijkl}$ is $\mathcal{R}^{-2}\equiv\mathcal{R}_{ij}^{ij}$ where $\mathcal{R}$ can be thought of as the radius of curvature of the space. The natural generalization of $\nabla^2\phi\propto\rho$ is given by $\mathcal{R}^{-2}\propto\rho=T_{ab}u^au^b$. Working out the left hand side and fixing the proportionality constant from the Newtonian limit, one finds that $G_{ab}u^au^b=8\pi T_{ab}u^au^b$. If this should hold for all observers (general covariance) then we need $G_{ab}=8\pi T_{ab}$ 
which is the standard result. But demanding that
 the equation  $\mathcal{R}^{-2}=8\pi\rho$ holds for each observer, captures the geometric statement --- viz. that energy density curves space ---  in a nicer manner. And it is the most natural generalization of $\nabla^2\phi\propto\rho$.}

GR makes several concrete predictions, the verification of which have bolstered our confidence in the theory over decades. Amongst many effects, I will describe three, of which  two are of historical importance.  The first effect deals with the orbits of planets around the Sun, which is most pronounced in the planet closest to the Sun, viz., Mercury. In  Newtonian theory, the orbit of Mercury would have been a closed ellipse with the Sun at one of the foci, if we treat it as an idealized central force problem. Astronomical observations have, however, shown 
that this is not the case and the direction of the major axis of the ellipse precesses by about 575 seconds of arc per century.
This, by itself, is no big deal because the perturbation due to other planets --- notably Jupiter and Saturn --- does cause this effect, but when  computed it leads to about 532 seconds of arc per century, leading to a discrepancy of about 43 seconds of arc per century. Based on the historical precedence --- which had actually led to the discoveries of Uranus and Neptune --- astronomers looked in vain for a planet Vulcan close to the Sun causing this discrepancy. The effects of GR precisely predicted an extra precession of 43 seconds of arc per century, thereby settling the issue.\endnote{Einstein used this fact to benchmark his field equations. It is  ironical that finally he obtained the $42''$ precession with the wrong field equation (which fortunately did not matter in the empty space limit) and working in the first order approximation for the metric. The latter was necessitated by Einstein's belief that the field equations are too complicated to yield an exact solution, because of the nonlinearity. Incredibly enough, if the spacetime is spherically symmetric and static, and is represented by the metric
\begin{equation}
ds^2=-(1+2\Phi)dt^2+\frac{dr^2}{(1+2\Phi)}+r^2(d\theta^2+\sin^2\theta d\phi^2)                                                                              \end{equation} 
then \textit{Einstein's equations become linear} with respect to the source $T^0_0=T^r_r\equiv\rho(r);\ T^\theta_\theta=T^\phi_\phi\equiv\mu(r)$. The general solution can be written  as (see e.g., p.302 of \cite{key7})
\begin{equation}
\Phi=-\frac{\alpha}{r}+\frac{G}{r}\int 4\pi r^2\rho(r) dr                                                         \end{equation} 
with  $\mu=\rho+ (1/2) r\rho'(r)$. So one can superpose solutions for $\rho=\rho_1+\rho_2+....$ etc to get $\Phi=\Phi_1+\Phi_2+.....$! This structural simplicity is one reason why Schwarzschild  discovered his metric within months of Einstein's field equations being  written down. (It took decades to understand the physics behind the Schwarzschild metric, but that is another story.)}

The second effect, which is probably more important both historically and astronomically, is the fact that gravity bends the paths of light rays.  This bending due to the Sun will make the apparent position of a star deviate from the actual position of the star by an angle which is of the order of arc-seconds. The effect will be maximum for a light ray coming with a grazing incidence on the solar rim, and is about 1.75 arc-seconds, which is precisely twice the prediction from  Newtonian  gravity if we thought of light as being made of particles moving with speed $c$. 
Obviously, such a deviation can be measured only during the total solar eclipse, when the stars around the Sun can be photographed.
Eddington led a pioneering expedition \cite{key1} to make this measurement on May 29, 1919, which dramatically verified Einstein's theory.  This made headlines around the world and made Einstein famous overnight, which remains somewhat puzzling to social scientists till today!

In the early days of relativity, this was considered a rather small deflection. We have come a long way since then, and the bending of the cosmic light rays by gravitational bodies,  even the Sun, is a \textit{huge} effect in today's astronomy in which stellar positions are known to a few milli-arc-seconds accuracy, and radio astronomers often talk about even micro-arcsecond measurements. What is more, the bending of light from cosmic sources leads to the phenomenon of gravitational lensing,\endnote{Though Einstein worked out the basics of gravitational lensing (by stars) in 1936, he concluded that: 
``It is of little value''. In contrast, Zwicky studied the lensing by galaxies, and in a truly visionary paper less than 1 page long (published in 1937), outlined all the diagnostic values of gravitational lensing --- testing relativity, magnifying faint objects, and measuring masses!} 
one of the most elegant and powerful diagnostic tools in astronomy today to measure the gravitational potential in the universe. 

Incidentally, this is not the only case where an effect of GR which was considered to be ``very small'' in the early days, turns out to be of great practical significance. The good quality GPS which you use routinely today, will become useless in a short span of time if the GPS satellites did not incorporate the effect of the gravitational field on the flow of time; it is routinely done in order for you not to lose your way while driving --- I can't think of a more practical use for a theory which was once considered rather abstract!

The third is the prediction of the existence of gravitational waves (the emission of) which have been detected  by R. A. Hulse  and J.Taylor in a beautiful series of observations  spanning over 30 years \cite{key2}. Since, in GR, gravitational effects propagate with the  speed as light, it also  contains gravitational wave solutions, just as Maxwell's equations possess electromagnetic wave solutions.\endnote{Again Einstein got it wrong at first! Nathan Rosen and Einstein wrote a paper claiming that gravitational waves do not exist, and sent it for publication to the Physical Review. The referee, distinguished cosmologist H.P. Robertson, discovered the error in the paper and, of course, rejected it. Einstein was sufficiently annoyed to write to the editor of the Physical Review, ``... on the basis of this incident, I prefer to publish the paper elsewhere ...'' and, in fact, never sent any of his subsequent papers to the Physical Review. Einstein resubmitted his (wrong) paper to the \textit{Journal of the Franklin Institute}  but --- fortunately, before it got published --- realized the mistake, informed the editor and finally published the correct version. (According to Leopold Infeld, who was Einstein's assistant, Robertson caught hold of Infeld and explained to him the error as well as the procedure to correct it, which Infeld conveyed to Einstein.) Gravitational waves do arise in Einstein's GR!}   
Cosmic sources, especially binary neutron stars orbiting each other, will emit such gravitational waves. This, in turn,  will make the binary system lose its energy, etc., causing the orbits of the stars to change in a clearly predictable manner. By monitoring the orbital parameters of such a system, we can verify that  gravitational waves exist and are emitted by such systems. This was done successfully by  Hulse and  Taylor, using the binary pulsar PSR B1913+16. The agreement with GR is  so remarkable  that there is no room for doubt about the existence of gravitational waves and their emission by this system.  It would be fun to detect the gravitational waves directly in the lab as well, but the experiments done over decades have failed due to technological limitations. Many are still underway, but contrary to what is sometimes portrayed,  these experiments are not necessary to verify the prediction of gravitational waves in Einstein's theory. This has already been done with exquisite detail, leading to Hulse and Taylor getting the Nobel Prize in 1993. 

GR has also contributed brilliantly to the way we understand several cosmic phenomena and let me describe a couple of them.

The first is the recognition of the astrophysical significance of black holes. Originally, black holes arose as rather esoteric solutions of Einstein's equations and --- in the initial years --- many leading physicists had difficulty in understanding and accepting such solutions as ``real physics''. Over decades, observational evidence has mounted for stellar mass black holes forming at the end stage of stellar collapse as well as for  supermassive black holes --- which are ubiquitous and exist in the centers of most galaxies. (Our own Milky Way  harbours such a massive black hole.) The  accepted explanation for  a class of objects ---  called active galactic nuclei (AGN) --- uses the concept of a black hole with an accretion disc around it. These AGN can be as small as  a  star and emit the energy equivalent to a galaxy  made of a hundred billion stars. Such super-luminous compact objects, when  first discovered, were a source of mystery; but today, they are routinely studied based on the paradigm of a black hole accreting matter from a surrounding disc. One could say that the entire branch of study based on AGN is propelled by the notion of black holes, which in turn is a child of GR. 

The  most significant feature of GR, however, is that Einstein's equations are capable of predicting the expansion of the universe. It would have been a great moment for human civilization if someone wrote down a couple of equations in a paper, solved them and predicted that the universe is expanding. But alas, this opportunity was missed because Einstein got cold feet and tinkered with his equations to make the universe static (by adding a term called the cosmological constant, which we will come across again). So, the expansion of the universe had to come from actual --- and fairly imprecise --- measurements \cite{key3}
 of distant galaxies  and only later had to be reconciled with the theory. Of course, the blame rests with good old Albie and not with GR. (As usual in physics, the correct equations are much wiser than their creators and if the creators don't second-guess the equations, everything will be fine!) 
Since then, we have come a long way in cosmology. Starting out as speculative and philosophical with  inaccurate observations and theoretical prejudices, it has evolved in the last 25 years or so to a precision science, mostly thanks to developments in technology. Today, observations lead the theory in cosmology and we theoreticians are trying hard to make sense of what the observations tell us.

Given all these feathers in its cap, can we consider GR to be the ultimate description of gravity, an epitome of perfect theory? \textit{Fortunately, no!} There is a lot of scope for improving on it, in spite of the  the gorgeous elegance it possesses. There are at least three theoretical aspects of GR which cry out for an extension or modification of the theory, in order for it to be considered satisfactory. Let me describe these open issues briefly. 

The first and foremost problem --- which I  consider to be \textit{the} problem in GR --- has to do with its lack of predictability in certain well-defined situations. Consider, for example, a very massive star which is collapsing to form a black hole. Prompted by scientific curiosity, one of your colleagues is willing to sit on the surface of the star and collapse with it, knowing fully well that she will not be able to communicate with you anything she finds. But she would like to know, before she starts on the trip, what fate awaits her. So she asks you --- the leading general relativist of the world --- to tell her what she could expect to see when the clock she carries with her shows different readings, 10 minutes, 10.1 minutes, ..... etc.  Your embarrassment is acute. \textit{Today, no theoretician in the world can answer her question.} They all have to say something like this: ``Well, when your clock shows a lapse of about 11.8 minutes, you will feel an extraordinarily strong gravitational field, and when it shows a lapse of 11.9343 minutes, a number I can compute to arbitrary precision, the gravitational field will become infinitely large. I regret I cannot predict  what happens at this moment or thereafter.'' 

The primary requirement of physical theories  is to predict the future evolution of a given dynamical system for an \textit{arbitrarily large} lapse of time, as measured by the clock carried by any observer; and GR categorically flunks this test. Mathematically, the matter in such a collapsing body hits a singularity (of infinite curvature and density and zero size) within a \textit{finite} time, as shown by the clocks moving with the collapsing matter ---  which is unacceptable in any sensible physical theory. Physically, one expects quantum gravitational effects to kick in when the sizes are of the order of the so called Planck length $L_P=(G\hbar/c^3)^{1/2}\approx 10^{-33}$ cm, built from the fundamental constants. But we do not know how to compute these quantum gravitational effects. \textit{All models of quantum gravity to date --- including those which their proponents want us to take seriously --- are  silent regarding this question, and have no rigorous, satisfactory, solution to offer.} This makes it the number one problem with GR.

The second theoretical conundrum about GR is related to what is usually called the cosmological constant problem --- except that it  actually arises from a serious structural defect of Einstein's theory and is a problem with gravity.  To appreciate this issue in its proper context, let us recall that, in all of non-gravitational physics, the  ``zero level'' of energy does not matter. What physical systems care about is the \textit{difference} in the energy between two configurations, rather than their absolute values. Mathematically, this arises from the fact that the dynamical equations describing a system do not change if you add a constant to the Hamiltonian, $H$, of the system; that is, the transformation $H \to H+C$, where $C$ is a constant, is a symmetry of the dynamical equations which describe all (experimentally verified) physical theories except GR. \textit{It turns out that Einstein's equations describing gravity break this symmetry!} You would have thought that elegant and beautiful theories should introduce a higher level of symmetry rather than go around breaking previously known exact symmetries; but this is precisely what happens. Any constant you add to your Hamiltonian will change the zero level of the energy, and in GR, gravity couples to this zero  level. This leads to two difficulties. 

First, we have reasons to believe that the zero level of the energy of the matter fields --- which act as the source of gravity --- has changed  during the evolution of the universe by fairly large amounts. Nevertheless, the cosmic gravitational field does not seem to have been affected by such changes. This requires either an extreme fine tuning of completely different sectors of physical theories --- which appears to be unnatural and unmotivated --- or the theory, as formulated, must be wrong. It should be possible to reformulate Einstein's theory in such a way that it respects the symmetry $H\to H+C$ and still leads to all the standard results. (I will say more about this later on.) 

The second issue concerns the actual numerical value of the zero level of the energy, which can be measured through its gravitational effects felt at very large cosmic scales. This contribution is expressed in terms of a parameter called the Cosmological Constant (and denoted by the symbol $\Lambda$) and has the effect of accelerating the expansion of the universe. (This cosmological constant is more popularly known as dark energy in the literature. But all observational evidence is consistent with dark energy being the cosmological constant, and hence I will call a spade a spade.) In standard units, $\Lambda$ has  the dimensions of the inverse square of length, and \textit{classical} gravity will be described by the dimension-full constants $G, c $ and $\Lambda$. It is silly to worry about the numerical value of a dimension full constant like $\Lambda$ and you cannot build a dimensionless number just from $G, c$ and $\Lambda$. The situation changes drastically when you accept that nature is quantum mechanical and we also have the Planck constant $\hbar$ at our disposal. It is now possible to construct a dimensionless number out of these four, viz., $\Lambda(G\hbar/c^3)=\Lambda L_P^2$. Observations tell us that this number has an incredibly tiny value of about $10^{-122}$! Our universe today has about 70 percent of the energy density contributing to its expansion coming from the zero level of energy (viz., $\Lambda$) which leads to this tiny but non-zero dimensionless number. Explaining this numerical value is the greatest challenge faced by theoretical physics today, and the conventional formulation of GR offers us no clue. 

The third theoretical issue with GR is probably the most tantalizing. It turns out that there is a peculiar relationship between gravity and thermodynamics which is universal in a manner which I will now explain. Recall that, because nothing can travel faster than light,  the paths of light rays in a spacetime determine which regions of spacetime can send and receive signals from which other regions. In the presence of gravity, the  light rays get bent and hence \textit{gravity} now determines  which regions of spacetime can communicate with which other regions. Further, the principle of equivalence --- which locally equates gravity to an accelerated frame --- demands that we treat all observers, inertial or accelerated, in a democratic way. It turns out that in any spacetime (even in flat spacetime) you can find observers who will not be able to receive signals from certain regions of spacetime; that is, these observers perceive a horizon in the spacetime beyond which they cannot see. (The black hole is the well-known illustration of this phenomenon, with respect to the observers who stay outside it, but this is just a specific example of a very general feature.) 

When you develop the quantum theory of other fields in such a spacetime, you are led to what I consider to be the most beautiful result we have in this subject: Any observer who perceives a horizon will attribute to it  a temperature 
\begin{equation}
T = (\hbar/k_B c)(\kappa/2\pi)\propto \kappa
\end{equation}  
where $\kappa$ is the acceleration of the observer \cite{key4}. 
This result makes the notion of temperature,  and, along with it, all of thermodynamics, an \textit{observer dependent} phenomenon. Consider, for example, the vacuum state of a field to which an inertial observer attributes zero temperature. An  accelerating observer  will see the same quantum state as a thermal state with the temperature proportional to her acceleration! You can no longer say ``this glass of water is at a temperature $T= 312$ K'' without specifying the  observer who is measuring it! When an inertial observer attributes the temperature  of 312 K to a glass of water, an accelerated observer will attribute to it a higher temperature. In fact, the notion of a particle itself becomes observer dependent when we have to take into account non-inertial observers.

So horizons lead to the (observer dependent) heating up of the spacetime.
Further, since horizons block information --- and lack of information is intimately related to entropy --- it is probably less surprising that observers will also attribute some amount of entropy to the horizons they perceive. 
It turns out that these thermodynamic features arise in a wide class of theories of gravity, much more general than Einstein's theory. All of them attribute the same temperature to the horizon but different entropies. In Einstein's theory, the horizon entropy is proportional to the area of the blocking surface, or, in other words, the entropy per unit area is a numerical constant. In other theories, the same spacetime horizon will be attributed a different entropy. This is again similar to the fact that, while you can heat a metal rod and a glass of water to the same temperature, the entropies they will have at that temperature will depend on their dynamical characteristics. You can keep two spacetime horizons at the same temperature, but the entropies they will have will depend on the dynamical equations which determine the spacetime structure. 

All this is very puzzling, as you would readily agree. Why should gravity have anything to do with thermodynamics and why should these relationships be so universal and transcend Einstein's theory? Of the three theoretical issues I have mentioned, this last one is the most fundamental, and possibly the most promising one to tackle. 

What does the future hold for GR, especially vis-a-vis the open issues like the ones mentioned above? You might have noticed that all these three issues  involve the Planck constant $\hbar$, and hence tackling them will require combining the principles of GR and quantum theory in a consistent manner. The most straightforward approach for constructing a quantum field theory --- which has been so successful in producing quantum electrodynamics and the electro-weak unification, leading to what is  called the standard model in particle physics --- relies on using a systematic \textit{perturbative} approach to construct and interpret the theory. Roughly speaking, this approach produces verifiable predictions from the theory by treating the interactions in a perturbative expansion in some small parameter in the theory. Further, it uses a specific technique (called perturbative renormalization) in order to give sensible meanings to divergent quantities which arise in the theory. While the proper interpretation of quantum field theory --- known as the Wilsonian approach --- demystifies all these at a conceptual level, we still do not know how to make predictions in any realistic quantum field theory if it is not perturbatively renormalizable. 

The trouble with gravity is that it is not perturbatively renormalizable for a large class of reasonable interactions. What is more, \textit{ every interesting question to which we want an answer from quantum gravity, is likely to be non-perturbative} in character. Since we do not know how to handle even quantum electrodynamics non-perturbatively with any level of generality, there is little hope that similar techniques will bear fruit in quantizing gravity.  This fact gained reluctant acceptance  rather slowly (circa the latter half of the 80s) among the high energy physicists. The last three decades witnessed significantly different and more imaginative approaches towards quantum gravity, but unfortunately --- often after a considerable amount of hope and hype --- none of them have led us to anywhere near answering the really important issues of quantum gravity. (Nevertheless, given the emotional investment of a generation of very talented physicists, the hope and the hype will continue!)

Given this backdrop, one might suspect that  we might have traveled a long way in the wrong direction as regards the interpretation of gravity and we need yet another paradigm shift.  One such approach --- which I am personally hopeful about --- is known as the emergent gravity paradigm \cite{key5}.  This approach takes the cue from  Boltzmann who told us: ``If you can heat it, it must have microstructure''. This allowed Boltzmann to interpret thermal phenomena in terms of of the (statistical) mechanics of the underlying discrete structures in matter, viz., the atoms and molecules. The smooth continuity of fluids and iron rods is an illusion valid at large scales when we average over the underlying discrete structures, but the latter manifests itself as the thermal energy of fluids and iron rods \textit{at the macroscopic scales}. Since we now know that spacetime can also be hot and possess entropy, it appears reasonable to study  the dynamics of spacetime exactly the way physicists studied matter before they knew what it was made of. 

This approach has been remarkably successful in several ways. To begin with, it allows one to obtain the dynamical equations of gravity from a thermodynamic extremum principle. The resulting equations restore to gravity the symmetry under the shifting of the zero level of energy. What is more, the cosmological constant arises as an integration constant to the solutions of the field equations, and its value can be fixed using an additional conservation law which the emergent paradigm attributes to our universe. This approach \textit{actually predicts \cite{key6}  the tiny numerical value of the cosmological constant} in terms of two other standard parameters in cosmology! 
One can show that:
\begin{equation}
\rho_\Lambda=\frac{4}{27}\ \frac{\rho_{inf}^{3/2}}{\rho_{eq}^{1/2}}\ \exp(-36\pi^2) 
\label{imp1}                                                                                  
\end{equation} 
where $\rho_\Lambda$ is the energy density contributed by the cosmological constant, $\rho_{inf}$ is the energy density during the inflationary phase and $\rho_{eq}$ is the energy density of the  radiation when the radiation and matter energy densities were equal. The three constant densities $(\rho_\Lambda, \rho_{inf}, \rho_{eq})$ are the signatures of our universe and are unrelated to one another  in standard cosmology; but the emergent paradigm connects the three! Eventually, high energy physics will determine the values of $\rho_{inf}$ and $\rho_{eq}$, allowing us to determine $\rho_\Lambda$ using \eq{imp1}. But at present, cosmological observations have precisely determined $\rho_{\rm eq}=  (\rho_{\rm matt}^4/\rho_{\rm rad}^3)
=[(0.86\pm 0.09) \ \text{eV} ]^4$ and $\rho_\Lambda=[(2.26\pm 0.05)\times 10^{-3}\text{eV}]^4$ using which one can  \textit{predict} that the inflationary scale to be $\rho_{inf}=(1-6)\times 10^{15}$ GeV. 
This prediction has \textit{verifiable consequences}, making the theory observationally disprovable --- which is more than  one can say about many other approaches. 

Further, it allows us to reinterpret much of classical gravity in a thermodynamic language, thereby demystifying the connection between thermodynamics and gravity. In fact, in this approach gravity \textit{is} the thermodynamics of the atoms of spacetime! It also explains why a very large class of theories share such a feature: The reason is same as why thermodynamics is applicable to a wide variety of physical systems, whether it is an iron rod or ionized plasma. Finally, the approach identifies the correct set of variables to describe gravity --- which happens to be quite different from what we use in the standard Einstein theory. So, by and large, this approach is successful in tackling two out of the three issues I have mentioned above. The issue of the singularity still needs to be addressed properly.

The main task ahead is to develop a fully microscopic theory of the ``atoms of spacetime'' and obtain the emergent gravity paradigm as its limiting case. It could very well be that such a microscopic theory is closely related to some of the existing candidate models for quantum gravity, possibly after some reformulation. For example, the ``top-down'' approach of the emergent gravity paradigm --- which is analogous to discovering the molecules from the thermal phenomena of matter --- leads to a holographic correspondence between (suitably defined) degrees of freedom living (i) in the surface and (ii) in the bulk region of spacetime. This is reminiscent of the notions of holographic correspondence  that arises in string theories, which are of course ``bottom-up'' models --- analogous to obtaining the thermodynamics  from  statistical mechanics. A marriage of the concepts in these two approaches could illuminate further the physical structure of both. 

Any such successful union will have deep implications about the very early stages of the universe, allowing us to eventually answer precisely the question which has always intrigued humanity: \textit{How did it all begin?} Whatever the answer is, it would be appropriate to consider it as a legacy of Einstein's genius.

\theendnotes

 \end{document}